\documentclass[
    ,final            % use final for the camera ready runs
  ]
  {aipproc}

\layoutstyle{6x9}

\newcommand{\be}{\begin{equation}}
\newcommand{\ee}{\end{equation}}
\newcommand{\bea}{\begin{eqnarray}}
\newcommand{\eea}{\end{eqnarray}}
\newcommand{\ceqn}[1]{equation~(\ref{#1})}

\newcommand{\like}{{\cal L}}
\newcommand{\mth}{m_{\rm th}}
\newcommand{\nth}{n_{\rm th}}
\newcommand{\nod}{{\cal N}}
\newcommand{\detect}{{\cal D}}

\begin{document}

\title{Accounting for Source Uncertainties in Analyses
of Astronomical Survey Data}

\author{Thomas J. Loredo}{
  address={Dept.\ of Astronomy, Cornell University}
}

\begin{abstract}
I discuss an issue arising in analyzing data from astronomical
surveys:  accounting for measurement uncertainties in the properties of
individual sources detected in a survey when making inferences about
the entire population of sources. Source uncertainties require the
analyst to introduce unknown ``incidental'' parameters
for each source.  The number of parameters thus grows with the size of
the sample, and standard theorems guaranteeing asymptotic convergence of
maximum likelihood estimates fail in such settings.  
From the Bayesian point of view, the missing
ingredient in such analyses is accounting for the volume 
in the incidental parameter space via marginalization.  I
use simple simulations, motivated by modeling the
distribution of trans-Neptunian objects surveyed in the outer solar
system, to study the effects of source uncertainties on inferences.
The simulations show that current non-Bayesian methods for handling
source uncertainties (ignoring them, or using an ad hoc incidental
parameter integration) produce incorrect inferences, with errors that
grow more severe with increasing sample size.  In contrast, accounting
for source uncertainty via marginalization leads to sound
inferences for any sample size.
\end{abstract}

\maketitle

%%%%%%%%%%%%%%%%%%%%%%%%%%%%%%%%%%%%%%%%%%%%
%% MAINMATTER
%%%%%%%%%%%%%%%%%%%%%%%%%%%%%%%%%%%%%%%%%%%%

\section{Introduction}

Astronomers devote enormous community resources to surveys: systematic
searches of some region of the sky, with
goals including characterization of populations of known astronomical
sources, and discovery of new sources.
Surveys play pivotal roles in nearly every astronomical discipline,
spanning the full range of scales from solar system astronomy (e.g.,
surveys of the asteroid and trans-Neptunian object (TNO) populations)
to cosmology (e.g., surveys of distant galaxies, active galaxies, and
cosmological gamma-ray bursts (GRBs)). Accurate and thorough analysis
of survey data is crucial to maximize the scientific return
from the extensive resources devoted to surveys.  But
although there is a high degree of sophistication in survey analysis
methods in isolated astronomical disciplines, in many disciplines more
rudimentary methods are used that waste information in the data and in
some cases can produce misleading conclusions.  Even in areas where
sophisticated methods are used, there is ongoing research in analysis
methods.
%methods, recognizing room for improvement in survey analysis
%methodology.

Several important and sometimes subtle issues arise in making accurate
inferences from survey data.  Here I will focus on one such issue: {\em properly accounting for individual source uncertainties}.
In the next section I briefly discuss how source uncertainties complicate
survey analysis by distorting the underlying distribution, emphasizing
that the distortions must be explicitly accounted for even (and
perhaps especially) when the number of data is large.  In \S~3, 
I present a concrete example illustrating the Bayesian approach
to handling source uncertainties---analysis of the magnitude distribution
of TNOs---including a comparison with results from non-Bayesian methods.

\section{Implications of Source Uncertainties}

Astronomers have known since the early 20th century that source
uncertainties distort the distribution of observables.  The clearest
early description of the effect is due to Jeffreys in 1938 \cite{Jeffreys38},
and is depicted in Fig.~1.  We measure and bin in equal ranges some
observable, say source distances, $r$.  Measurement errors produce estimates $\hat r$ that
differ from the true values, so some measurements will be put in the
wrong bin.  If the true number in a bin is greater than that in its
neighbors, we expect more measurements to be scattered out of the
bin than into it.  As a result, the overall distribution is distorted;
it gets smoothed in a manner resembling a convolution.

\begin{figure*}[t]
\centering
\includegraphics[width=30pc]{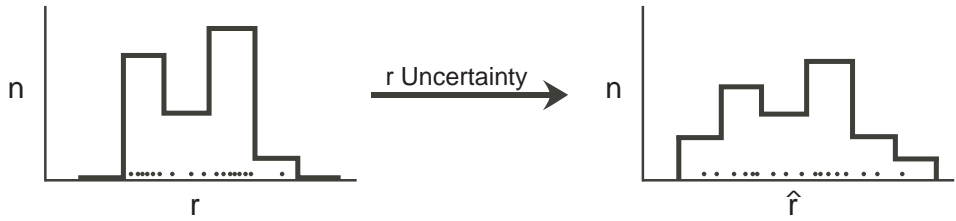}
\caption{Distortion of the source observable distribution due
to source uncertainties.}
\end{figure*}

Can the distortion be removed? Jeffreys' brief paper criticized a
solution offered by Eddington that treated the problem as one of
inverting a convolution; Eddington's solution, predating understanding
of the ill-posed nature of such an inversion, was highly unstable.
Jeffreys pointed out that a superior approach would be to {\em predict}
rather than invert the data; he suggested introducing a parameterized
model and using the likelihood function to find the model parameters
that best predict the data.  But he did not present such a solution in
any detail, and his advice was largely ignored until the 1980s.

In the meantime, Malmquist offered an approach whose basic
features have guided survey analyses to this day.  Rather than use
the naive best-fit estimates of $r$, Malmquist showed that one
could use the underlying distribution, $f(r)$, and the known
size of the $\hat r$ uncertainties to devise corrections that
lead to revised estimates, $\tilde r$, that can be used
to estimate $f(r)$ in an unbiased manner.  Malmquist's work
was so influential that biases resulting from ignoring
source uncertainty have come to be known as 
{\em Malmquist biases}.\footnote{See Strauss \& Willick (1995), 
Teerikorpi (1997) and Sandage and Saha (2002) for some recent reviews of
Malmquist-type survey biases.  Lutz-Kelker bias is a similar bias arising due to source
uncertainties.  For brevity's sake we do not distinguish it from Malmquist
bias here; see Smith (2003) for brief remarks on the relationship.}

A serious defect of Malmquist's approach is that the corrections depend
on $f(r)$, which is unknown and often the object of interest in the
investigation.  Malmquist was interested in estimating the density of
stars in space.  He assumed that the {\em volume} density is uniform,
so that the $r$ distribution is nonuniform, with $f(r)\propto r^2$ (due
to the $r^2$ growth of the spherical coordinate volume element).  This
assumption was used for many years, but in many applications it is
unsound. This is particularly the case in studies of galaxy surveys,
since we know that the galaxy distribution is very nonuniform. As a
result, generalizations of Malmquist's approach have been sought.  Some
assume a simple parameterized form for the density (typically a power
law) and determine corrections as a function of the parameter.  Some
procedure is then devised to set the parameter using the data, e.g.,
via an iterative scheme.  Other approaches have an empirical Bayesian
flavor (though they are typically labeled as maximum likelihood
approaches); they multiply the likelihood for the $r$ value of a source
by a prior determined by the source density.  For a uniform density (so the
prior is $\propto r^2$), taking $\tilde r$ equal to
the posterior mean value of $r$
duplicates Malmquist's corrections.  Again, the density can be
parameterized and the parameter adjusted using some criterion to
measure consistency between the final $r$ estimates and the prior. But
no rigorous, self-consistent approach has yet been offered, and
research continues on how best to account for Malmquist bias.

Nothing about Jeffreys' or Malmquist's observations depended on
the observable being distance, and the effect will be present for
any other observable with a nonuniform distribution that is measured 
with uncertainty.  Unfortunately, although the importance of correcting
for such distortion is well known for space density estimation, in other
applications the effect has not been so widely recognized.

The dependence of Malmquist corrections on the unknown density is a well known
problem with Malmquist's treatment of the effects of source uncertainty.  But
there is a more subtle potential problem that is not so widely recognized. The
Malmquist approach replaces naive point estimates of observables with
``corrected'' point estimates. Although source uncertainties are used to
determine the corrections, once the corrections are determined, the source
properties are treated as precisely known in subsequent analysis.  But
ignoring such uncertainties can be dangerous, as the following example
illustrates.

In a classic paper written in 1948 motivated by statistical problems in
astronomy, Neyman and Scott \cite{NS48}\ discussed the following problem. For
each of a number of sources we make repeated measurements of the source
intensity with an instrument that adds noise to the signal. The sources each
have different intensities.  We assign a Gaussian distribution for the noise,
but the instrument's noise standard deviation, $\sigma$, is not known
at the outset.  We can pool together the data from $N$ sources to estimate the
common parameter, $\sigma$ (and then use this information to 
estimate the source intensities).

As the simplest case, suppose there are two measurements for each object.  For
source $i$, the likelihood function for its intensity, $\mu_i$, and $\sigma$
is just the product of two Gaussians,
\be
\like(\mu_i,\sigma) =
  \frac{1}{\sigma\sqrt{2\pi}} \exp\left[-{(x_i-\mu_i)^2\over 2\sigma^2}\right] 
  \times \frac{1}{\sigma\sqrt{2\pi}} \exp\left[-{(y_i-\mu_i)^2\over 2\sigma^2}\right],
\ee
where $x_i$ and $y_i$ denote the two measurements. Fig.~2a shows
contours of this likelihood for 
typical measurements with true $\sigma=1$.  As one might expect, the
likelihood is symmetric in $\mu_i$ with its peak at the sample mean,
$(x_i+y_i)/2$.  The $\sigma$ uncertainty is large for just two
measurements, but includes the true value.  The dashed curve labeled
``1 pair'' in Fig.~2b shows a standard frequentist summary of the
implications of this data for $\sigma$. The curve is the {\it profile
likelihood} for $\sigma$, the maximum likelihood as a function of
$\sigma$ (i.e., maximized with respect to $\mu_i$).  Also shown is the
Bayesian summary of the information in the data about $\sigma$, the
marginal likelihood for $\sigma$, obtained by multiplying the
likelihood by a prior for $\mu_i$ (here uniform) and integrating out
$\mu_i$. (A final Bayesian inference for $\sigma$ would be found by
multiplying this by a prior density for $\sigma$ to get the marginal
posterior for $\sigma$.)  The marginal likelihood is different from the
profile likelihood, but not significantly so, given the large
uncertainties.

\begin{figure}[t]
% Put width first to avoid graphicx scale factor problem; after
% rotn it sets the *height*.
\centering
\includegraphics[width=2.in,angle=-90]{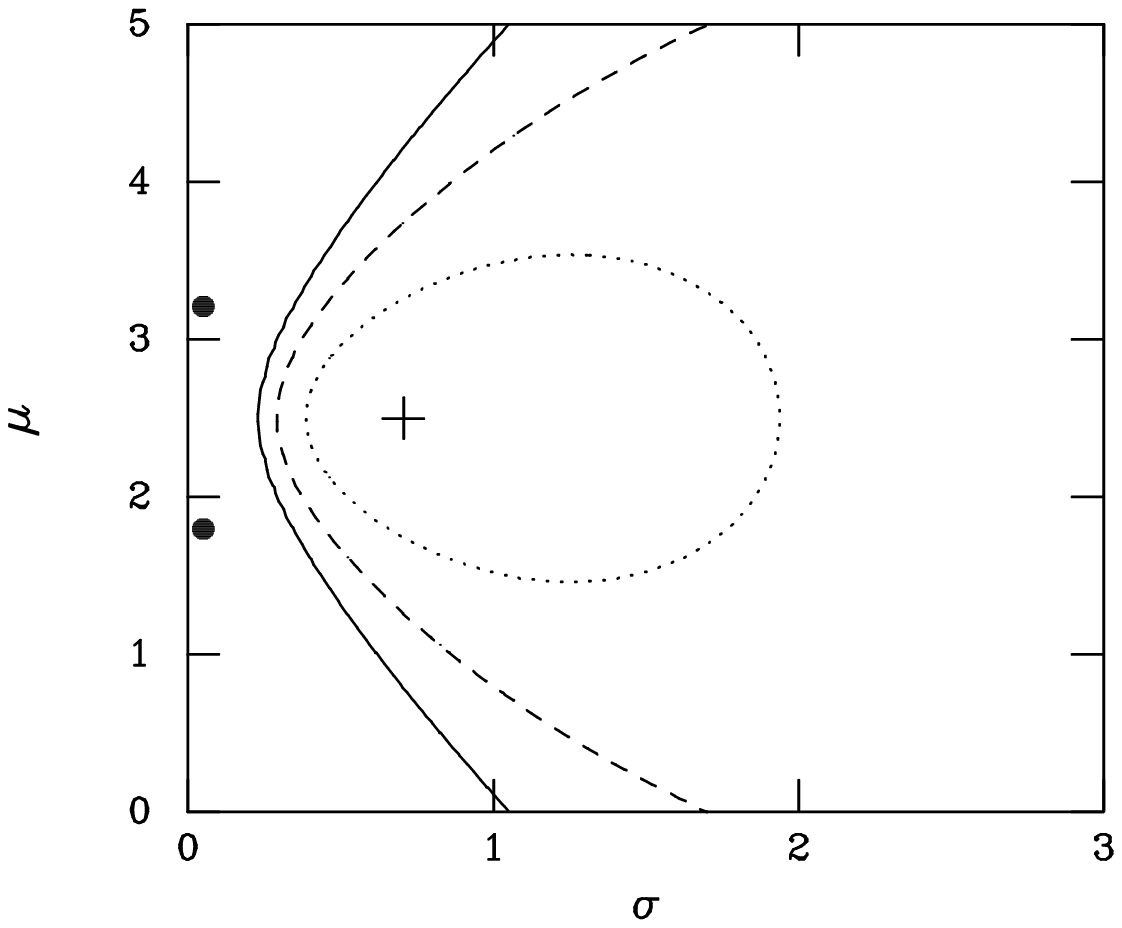}\qquad
\includegraphics[width=2.in,angle=-90]{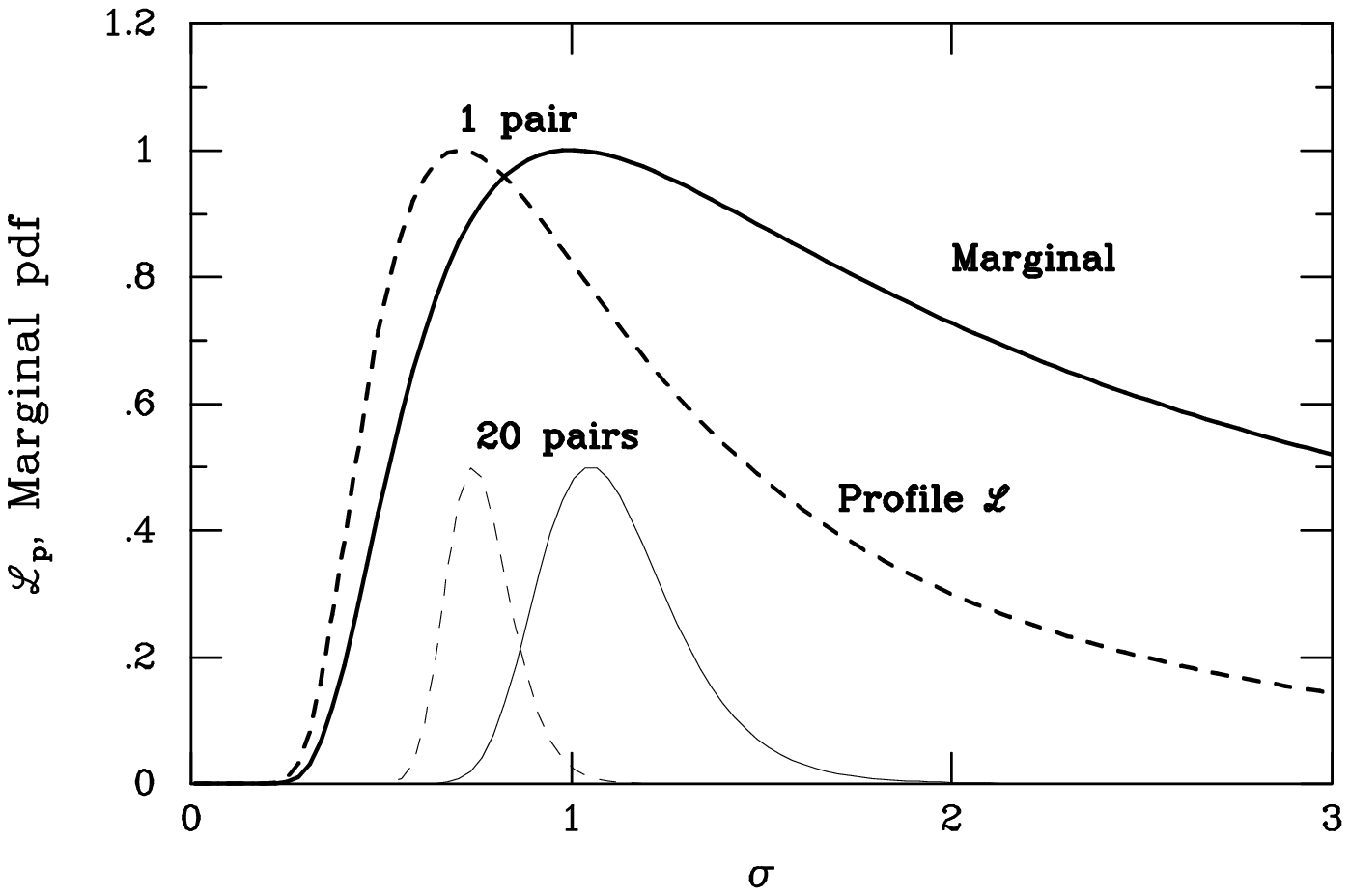}
\caption{The Neyman-Scott problem.  {\it (a)}, left, the likelihood
for $\mu$ and $\sigma$ for a single pair of measurements (with measured
values $x$ and $y$ shown by dots along the $\mu$ axis).  Cros shows
the maximum; contours are at the asymptotic 68\% (dotted), 95.4\% (dashed)
and 99.73\% (solid) confidence levels.
{\it (b)}, right, the profile likelihood (dashed) and marginal likelihood
(solid) for $\sigma$ for a single pair, and for 20 pairs (ordinates rescaled to
facilitate comparison).}
\end{figure}

Now consider what happens when we pool information from observations of 20
sources (by multiplying likelihoods).  The curves labeled ``20 pairs'' show
the results using Monte Carlo samples. The profile likelihood is converging
quickly to the {\it wrong} value (it is easy to show that it asymptotically
converges to $\sigma/\sqrt{2}$).  The marginal distribution converges less
quickly, but is peaked near the true value (and in fact is consistent---it
asymptotically converges to the true value for any smooth prior). Returning to
Fig.~2a, we can understand this behavior.  The likelihood contours for a
single pair of measurements are asymmetric in $\sigma$, enclosing much less
volume in the $\mu_i$ direction at small values of $\sigma$ than at larger
values.  Focusing on the likelihood {\it peak} ignores this, leading the
maximum likelihood approach astray, while accounting for the {\it volume}
under the likelihood via marginalization gives correct inferences.

This problem is notable in two respects.  First, it is tempting to hope
that as one gathers more and more data, asymptotics comes to the rescue
and guarantees convergence to the truth, with uncertainties ``averaging
out.''  The example shows this is not true.  The reason is that the the
presence of source uncertainties (here in $\mu_i$) means that each
source brings with it a new parameter that must be estimated,
explicitly or implicitly.  Neyman and Scott called these ``incidental''
parameters, in contrast to the ``structural'' parameter, $\sigma$, 
shared among all measurements.  The total number of parameters
present grows with the number of data so the allowed volume in parameter
space does not shrink to zero asymptotically, and this feature of such
problems prevents the usual asymptotic guarantees from holding.

Second, the example shows that even frequentist maximum likelihood methods,
close in many respects to Bayesian methods, are insufficient in such settings.
It is the Bayesian focus on {\em volumes} under likelihoods, a consequence of
taking a probabilistic approach to parameter uncertainty, that leads to
accurate inferences.

The Neyman-Scott problem is not merely academic.  One of the problems
that motivated their work frequently arises in astronomy and other disciplines:
fitting data that has errors in both the abscissa and
the ordinate (``errors-in-variables'' models).  In such problems, the
unknown true abscissa values appear as nuisance parameters, and their
number grows with the number of data.  As a result, one must handle
such problems with some care, a point made in these conferences by
Ed Jaynes and Steve Gull (see \cite{Gull89}).

Of course, another setting where the features of Neyman's and Scott's
problem appear is survey analysis; each source brings with it its own
uncertain incidental parameters, and we learn about shared structural
parameters describing the distribution of sources by pooling the
information from the sources.  To demonstrate the relevance of the
problem to survey analysis, let us consider a concrete example.

%======================================================================
\section{Trans-Neptunian Objects}

We consider surveys detecting and reporting apparent magnitudes of TNOs, a
large population of minor planets with orbits extending beyond that of
Neptune.  The first TNOs were detected in 1992; today nearly 1000 are known. 
Several dynamically distinct populations comprise TNOs, distinguished by the
distributions of their orbital elements.  The majority are ``classical''
Kuiper belt objects (KBOs), with a broad distribution of low-eccentricity,
low-inclination orbits. Other populations have eccentric orbits due to
interactions with Neptune (``scattered''
KBOs) or orbits in various resonances with Neptune (e.g., Plutinos, in orbits
with a $2:3$ resonance with Neptune; Pluto is considered a member of this TNO
class).  We focus on classical KBOs here; see Gladman et al.\ \cite{Getal01}\
for a brief overview of the populations, and the review article
by Luu and Jewitt \cite{LJ02}, discoverers of the first TNO, for a more
thorough overview. Of the $\sim 10^3$ known TNOs, only $\sim 10^2$ were
discovered in surveys that have been characterized sufficiently to
allow a rigorous analysis of the TNO distribution; this number will
grow rapidly in the next few years.  Fig.~3 presents a plan view of the
solar system showing the current locations of 200 of the earliest
discovered TNOs, and their relationship to the planets in the outer
solar system.

\begin{figure*}[t]
\centering
\includegraphics[height=2.2in]{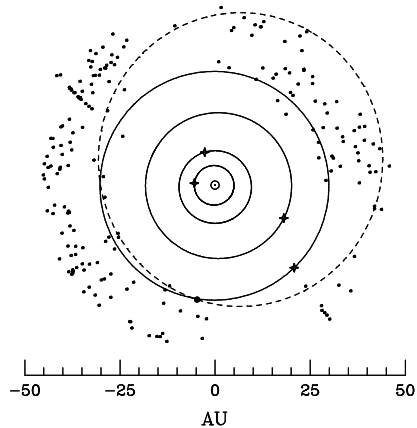}
\caption{A plan view of the outer solar system in July 2004, looking
down the north ecliptic pole.  The solid curves and open crosses show
the orbits and locations of Jupiter, Saturn, Uranus and Neptune;
$\odot$ denotes the Sun.  The dashed curve and large dot show the orbit
and location of Pluto, considered to be the largest of a class
of TNOs called Plutinos.  The small dots show the locations of
200 TNOs; about 800 have so far been discovered.}
\end{figure*}

Suppose TNOs have a distribution of sizes, $D$, that is a power law,
$f(D) \propto D^{-q}$, and a density distribution that varies with
heliocentric radius, $r$, as a power law, $n(r) \propto r^{-\beta}$,
bounded between Neptune's orbit and some maximum distance. TNOs
are seen by reflected sunlight, so the flux from a TNO obeys a
$F\propto D^2/ r^{4}$ law (ignoring for simplicity Earth's 1~AU
distance from the sun, small compared to TNO distances). 
The flux distribution of TNOs is then a broken power
law, with the power law index of dim objects determined by $\beta$,
and that of bright objects determined by $q$.  Astronomers
report optical fluxes on a (negative) logarithmic {\em magnitude}
scale, with magnitude $m = -2.5\log_{10} F/F_0$ (with $F_0$ a fiducial
flux).  Let $\Sigma(m)$ be the number of TNOs per square degree with
magnitudes less than $m$.  The broken power law flux distribution implies a
cumulative magnitude distribution that is a broken exponential,
\be
\Sigma(m) = 10^{\alpha (m-m_0)},
\ee
where the log-slope, $\alpha$, is different at large and small $m$.
For the dimmest objects (large $m$), the slope measures the TNO space
density index, with $\beta = 10\alpha + 3$.  For bright TNOs
(small $m$), the slope measures the size distribution index, with $q =
5\alpha + 1$.  Note that the parameters of physical interest, $q$ and
$\beta$, are related to the more directly observable slope $\alpha$ by
large factors so that small errors in $\alpha$ lead to large errors in
the inferred physics.  A goal of TNO survey analysis is to estimate the
slope of the magnitude distribution in different magnitude ranges in
order to estimate the indices of the size and density distributions.

To date well over a dozen well-characterized surveys of the TNO
population have been undertaken, with extremely diverse
characteristics.  Some spread observing resources in relatively short
exposures over broad regions of the sky along the ecliptic, providing
the best data on bright TNOs (since they sample a large area).  Others
focus all resources on a narrow ``pencil beam'' target area, using
repeated long exposures to search for the numerous very dim TNOs.  Many
surveys are unsuccessful in the sense of not detecting any new TNOs;
nevertheless the lack of detections is itself useful information that
sets bounds on the TNO density in the regions accessible to such 
surveys. Different surveys use filters that access different parts of
the optical spectrum.  An important challenge in TNO survey analysis is
how to consistently combine the information from the many surveys.

TNOs are dim objects that are challenging to detect and measure.  Measurement
uncertainties due to photon counting statistics can be significant. 
Systematic uncertainties, e.g., due to the need to convert measurements to a
common wavelength range, or due to TNO variability, can be significant.  Both
statistical and systematic uncertainties tend to be largest for the dimmest
TNOs.  For bright TNOs the magnitude uncertainty is relatively small, $\sim
5$\%.  For dim TNOs, it is typically much larger, 20--30\%.  Since distant TNOs
tend to be dim, and since the volume for a given radius range is larger for
distant sources than for nearby sources, many or most of the sources in a
particular survey tend to be dim.  Thus many of the detected TNOs have
significant magnitude uncertainties.  An additional challenge in TNO survey
analysis is accounting for these uncertainties.

Most studies of TNO data use approaches that can be characterized as
trying to ``fix the data.'' They try to construct a ``debiased''
estimate of $\Sigma(m)$ that has the selection effects of a particular
survey removed.  These estimates are then combined and models fitted
via least squares methods. These approaches have numerous defects.  The
data are sparse; bins must be wide to contain enough objects to justify
the asymptotic approximations underlying the methods.  This sacrifices
resolution; some studies simply ignore the requirement and allow bins
to have few counts rather than use wide bins.  The arbitrary
choice of bins introduces troubling subjectivity into the results.
Some studies fit the {\em cumulative} rather than differential binned
distribution, ignoring the strong correlations in the binned estimates
and thus seriously underestimating the uncertainty in the final
inferences.  None of the studies make any attempt to account for
magnitude uncertainties, even though they are typically of comparable
scale to the bin widths for dim sources.  With such uncertainties,
consistent model-independent ``debiasing'' is simply impossible;
moreover, ignoring them risks underestimating uncertainties in the
final inferences, and possibly finding incorrect results due to the
volume effects discussed above.

Bayesian inference is ideally suited to the challenges of TNO survey
analysis.  Information from disparate surveys can be easily
combined by multiplying likelihoods.  Source uncertainties can be
handled by introducing nuisance parameters and marginalizing.
Systematic error can be incorporated in the analysis, since the
Bayesian approach does not restrict use of probability distributions
only to ``random'' uncertainties.

Gladman et al.\ \cite{Getal98,Getal01}\ have adopted a Bayesian approach to
TNO survey analysis, building on the work of Loredo and Wasserman on analysis
of GRB survey data \cite{LW95,LW98a,LW98b}.  Their results differ markedly from
those of investigators using the approaches described above.
More recently, Bernstein et al.\ (\cite{Betal04}; B04) have advocated a 
quasi-Bayesian
approach, but with an incorrect likelihood function.   In the
remainder of this section, I will describe the Bayesian approach and how it
differs from the B04 approach, illustrating how some of the concerns of the
previous section can manifest themselves in analyses of TNO survey data.
The results show that it is not enough to follow the Bayesian approach
``in spirit;'' inferences can be significantly corrupted unless the
Bayesian prescription is followed with care.

For an analysis of the TNO magnitude distribution, the information from a TNO
survey can be summarized by reporting the following quantities:
\begin{itemize}
\item The solid angle, $\omega$, 
examined by the survey;
\item The survey efficiency function, $\eta(m)$,
specifying the probability that a TNO of magnitude $m$ will produce
data meeting the survey criteria for detection;
\item Source likelihood functions, $\ell_i(m)$, giving the likelihood
that TNO $i$ has magnitude $m$.  By definition $\ell_i(m) = p(d_i|m,M)$ is 
the probability for the data $d_i$ from source $i$ presuming
the source has magnitude $m$, with $M$
denoting any data modeling assumptions.  It will
often be adequately summarized by a Gaussian function specified
by the best-fit (maximum likelihood) $m$ value for the TNO and
its uncertainty.
\end{itemize}
Our goal here is to infer the parameters, $\theta$, of a specified
model for the TNO magnitude distribution.  The key ingredient in a
Bayesian analysis is the likelihood function for $\theta$ based on all
the survey data, $D$, defined by $\like(\theta)=p(D|\theta,M)$.  We
will derive it in two steps.  First, we derive the likelihood for an
idealized survey able to detect every TNO brighter than some magnitude,
$\mth$, and reporting the magnitude of TNO $i$ precisely as $m_i$.
Then we account for the complications of detection efficiency and
source uncertainty.

We model the magnitude distribution as a Poisson point process
specified by the differential magnitude distribution, $\sigma(m)$,
defined so that $\sigma(m)dm d\omega$ is the probability for there
being a TNO of magnitude in $[m,m+dm]$ in a small patch of the sky of
solid angle $d\omega$ (so $\Sigma(m)$ is its integral).  For idealized
data, we imagine the $m_i$ values spread out on the magnitude axis,
shown in Fig.~4.  We divide the axis into empty intervals indexed
by $\alpha$ with sizes
$\Delta_\alpha$, between small intervals of size $\delta m$ containing
the $N$ detected values $m_i$.  The expected number of TNOs in empty
interval $\alpha$ is,
\be
\mu_\alpha = \omega \int_{\Delta_\alpha} dm\; \sigma(m).
\label{mua-ideal}
\ee
The expected number in the interval $\delta m$ associated with
detected TNO $i$ is
\be
\mu_i = \omega\,\delta m\,\sigma(m),
\label{mui-def}
\ee
where we take $\delta m$ small enough so the integral over $\delta m$
is well approximated by this product.  The probability for seeing no
TNOs in empty interval $\alpha$ is the Poisson probability for no
events when $\mu_\alpha$ are expected, given by $e^{-\mu_\alpha}$.  The
probability for seeing a TNO of magnitude $m_i$ in $\delta m$ is the
Poisson probability for one event when $\mu_i$ are expected, given by
$\mu_i e^{-\mu_i}$.  Multiplying these probabilities gives the
likelihood for the parameters, $\theta$, specifying $\sigma(m)$.  The
expected values in the exponents sum to give the integral of
$\sigma(m)$ over all accessible $m$ values, so the likelihood can be
written,
\be
\like(\theta) = (\omega\delta m)^N \exp\left[-\omega\int dm\, \Theta(\mth-m)\sigma(m)\right]
\prod_{i=1}^N \sigma(m_i),
\label{L-ideal}
\ee
where $\Theta(\mth-m)$ is a Heaviside function restricting the integral
to $m$ values smaller than $\mth$.  The factor in front is a constant
that will drop out of Bayes's theorem and can henceforth be ignored.

\begin{figure*}[t]
\centering
\includegraphics[width=25pc]{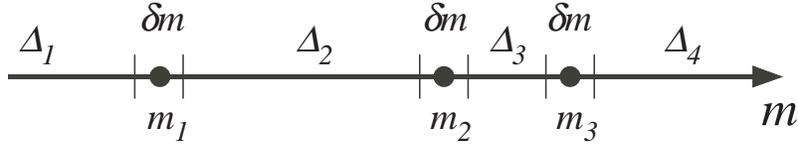}
\caption{Constructing the likelihood for an idealized survey as
a point process in $m$, with empty intervals of sizes $\Delta_\alpha$
and TNOs detected at points $m_i$ in small intervals of size $\delta m$.}
\end{figure*}

Now we consider the real survey data, which differs from the idealized
data in two ways:  the presence of a survey efficiency rather than a
sharp threshold, and the presence of magnitude uncertainties.  We
immediately run into difficulty with a point process model because we
cannot make the construction of Fig.~4, since we do not know the precise
values of the TNO magnitudes.  But in a Bayesian calculation we can
introduce these values as nuisance parameters, and then integrate them
out.  To facilitate the calculation, we need to introduce some
notation.   When occurring as an argument in a probability, let 
$m_i$ denote the proposition that there is a TNO of magnitude $m_i$
in an interval $\delta m$ at $m_i$.  We divide the data, $D$, into
two parts: the data from the detected objects, $\{d_i\}$, and the
proposition, $\nod$, asserting that no other objects were detected.
Then the likelihood can be written,
\bea
\like(\theta) 
  &=&  p(D|\theta,M)\nonumber\\
  &=& \int \{dm_i\}\; p(\{m_i\},\nod|\theta,M)\, p(\{d_i\}|\{m_i\},\nod,\theta,M).
\label{L-marg}
\eea

The first factor in the integrand can
be calculated using a construction similar to that used for the idealized
likelihood above, with one important difference: the presence of
the $\nod$ proposition means that we cannot assume that no TNO
is present in a $\Delta_\alpha$ interval, but rather that no TNO
was {\em detected}.  Thus
these probabilities are Poisson probabilities for no events when
$\mu_\alpha$ are expected, with
\be
\mu_\alpha = \omega \int dm\; \eta(m)\,\sigma(m),
\ee
the {\em detectable} number of TNOs in the interval rather than the
total number.  Thus the first factor in the integrand
in \ceqn{L-marg}\ resembles the right hand side of \ceqn{L-ideal},
but with $\eta(m)$ replacing the Heaviside function in the integral 
in the exponent.

The second factor in the integrand in \ceqn{L-marg}\ is the probability
for the data from the detected objects, given their magnitudes.  Since
the source likelihood function is by definition $\ell_i(m)=p(d_i|m)$,
this probability is just a product of source likelihood functions
evaluated at the specified $m_i$ values (with the $m_i$ values
given, the $\nod$ and $\theta$ propositions in this probability
are irrelevant to the probability for $d_i$).  Now we can 
calculate \ceqn{L-marg}:
\be
\like = \exp\left[-\omega\int dm\, \eta(m)\sigma(m)\right]
\prod_i \int dm\, \ell_i(m)\sigma(m),
\label{L-final}
\ee
where we have dropped the unimportant interval factors, and
we have simplified the notation by dropping the indices from
the $m_i$ variables in the integrals, since they are just
integration variables for independent integrals.

B04 derived a likelihood function for TNO magnitude data by an
informal argument, with their final result being,
\be
\like_B = \exp\left[-\omega\int dm\, \eta(m)\sigma(m)\right]
\prod_i \int dm\, \eta(m)\ell_i(m)\sigma(m).
\label{L-B}
\ee
This differs from \ceqn{L-final}\ in the presence of $\eta(m)$ factors
in the source integrals.  B04 argued that these factors should be
present because the probability for the data from a detected TNO
with given magnitude
should be the product of the probability that the TNO was detected,
given by $\eta(m)$, times the probability for the detection data.
But this is an incorrect calculation of a joint probability.\footnote{I made
the same error in an early analysis of the neutrinos detected from
SN~1987A \cite{LL89}.  In later work the error was corrected and
discussed along the lines presented here \cite{LL02}.}  Let
$\detect_i$ denote the proposition that a TNO is actually detected
in the data from TNO candidate $i$.  The product rule lets us
calculate the joint probability for $\detect_i$ and $d_i$ in two
ways:
\bea
p(\detect_i,d_i|m,M)
  &=& p(d_i|m,M) p(\detect_i|d_i,m,M)\nonumber\\
  &=& p(\detect_i|m,M) p(d_i|\detect_i,m,M).
\label{p-Dd}
\eea
In the first line, the first factor is just the source likelihood,
$\ell_i(m)$.  The second factor is the probability that TNO $i$
is detected, {\em given the data from that TNO}.  But since we
are considering data from a detected TNO, this probability is
unity by definition (i.e., detection is a criterion that the
observed data are in some acceptable set, and by definition data
from a detected TNO must lie in that acceptable set).  So the
joint probability for detection and the data, given $m$, is just the
isolated $\ell_i(m)$ term appearing in the correct likelihood.

Now examine the second line in \ceqn{p-Dd}, corresponding to the
factorization implicitly used by B04.  The first factor
is the probability that a TNO of magnitude $m$ would be detected;
this is given by $\eta(m)$.  But the second term is the probability
for the data from the detected TNO, {\em given that it has been
detected}.  We can calculate this probability with Bayes's theorem;
it is given by $\ell_i(m)/\eta(m)$.  The $\eta$ factors cancel, and
this factorization of the joint probability also equals $\ell_i(m)$,
as it must.  The B04 derivation fails to condition on detection when
calculating the data probability, and so produces an incorrect final
likelihood function.  

To make these considerations concrete, imagine a simple detector that
counts photons in a single pixel, and reports a detection if the counts
exceed some threshold, $\nth$.  The data from detected source $i$ is
just the counts, $n_i$, detected from that source.  Then the first
factorization is the product of the Poisson probability for $n_i$
counts given the magnitude, and the probability that $n_i>\nth$.  Since
the source was detected, the last probability is unity, and we are left
with the Poisson likelihood for $m$ defining $\ell_i(m)$.  For the second factorization, the
first factor is the probability that $n_i>\nth$ given the TNO
magnitude.  This is a sum of Poisson probabilities for counts above
$\nth$ (it is given by an incomplete Gamma function).  The second
factor is the probability for seeing $n_i$ counts from a source of
magnitude $m$, {\em given that the counts from that source are above
$\nth$}.  This is the Poisson probability for $n_i$, but renormalized
for counts above the threshold.  The renormalization requires division
by the sum given by the first factor, so that factor cancels and again
we are left with the Poisson probability for $n_i$ counts given $m$,
that is, $\ell_i(m)$ (with no $\eta(m)$ factor).

B04 further argued that the uncertainties were small enough that they
could be ignored, essentially taking $\ell_i(m)$ to be
a $\delta$-function at the best-fit magnitude, $\hat m_i$.  In this
approximation the integral for object $i$
becomes $\eta(\hat m_i) \sigma(\hat m_i)$.
The first factor is constant with respect to the model
parameters, so this corresponds to a likelihood function given by,
\be
\like'_B = \exp\left[-\omega\int dm\, \eta(m)\sigma(m)\right]
\prod_i \sigma(\hat m_i).
\label{L-Bp}
\ee
This is the idealized likelihood of \ceqn{L-ideal}, with the Heaviside
function replaced by the detection efficiency.  This is the likelihood
actually used by B04.

To explore the consequences of use of the incorrect likelihood of
\ceqn{L-B}, and of completely ignoring uncertainty and using \ceqn{L-Bp},
we can construct simple simulated surveys where the truth is known and
see if these likelihoods recover the truth.  I simulated data from
a TNO distribution with a ``rolling'' power law index using a model
advocated by B04,
\be
\sigma(m) = \sigma_{23} 10^{[\alpha(m-23) + \alpha'(m-23)^2]},
\ee
where $\sigma_{23}$ is the density at $m=23$, $\alpha$ is the power
law slope at $m=23$, and $\alpha'$ is the rate of change of the slope
with $m$.  TNO magnitudes were drawn from this distribution, and then
the counts expected from that source in a simple single-pixel measurement
were sampled from a Poisson distribution with expected value proportional
to the flux.  If the counts were above a threshold, $\nth$, the TNO was
detected, and its Poisson likelihood was used for $\ell(m)$.  For the
results shown in Fig.~5, the threshold and expected counts were chosen
so that the dimmest detected TNOs had an uncertainty $\sim 33$\%.  The
model used has $\alpha=0.75$ and $\alpha'=-0.05$, a model B04 find
describes classical KBOs well.  Fig.~5a shows the most probable
parameter values from 100 simulated surveys with $N=100$ detected TNOs;
the values were found by calculating the marginal distribution for
$\alpha$ and $\alpha'$ using flat priors.  The solid dots show
estimates using the correct likelihood; the open circles show estimates
ignoring source uncertainty, using \ceqn{L-Bp}.  Estimates from the
correct likelihood are scattered roughly symmetrically about the true
value (indicated by the open $\times$).  Estimates ignoring source
uncertainties systematically overestimate $\alpha$ and underestimate
$\alpha'$.  For samples of this size (comparable to the size of the
sample analyzed by B04), the uncertainties are large enough that the
estimates are still sometimes near the correct value despite the strong
bias.  Fig.~5b shows estimates with $N=1000$, and the situation is
worse----the estimates ignoring uncertainty are converging away from
the truth, similar to the behavior seen in the Neyman-Scott problem.
In contrast, the correct likelihood produces estimates converging on
the true value.  Figs.~5c,d repeat the experiment using the 
likelihood of \ceqn{L-B}\ that attempts to include uncertainties, but
has the incorrect $\eta(m)$ factor.  We find the same behavior,
indicating that even though \ceqn{L-B}\ has integrals over the source
uncertainties, the incorrect $\eta(m)$ factors corrupt inferences using
this likelihood.

\begin{figure*}[t]
\centering
\includegraphics[height=30pc,angle=-90]{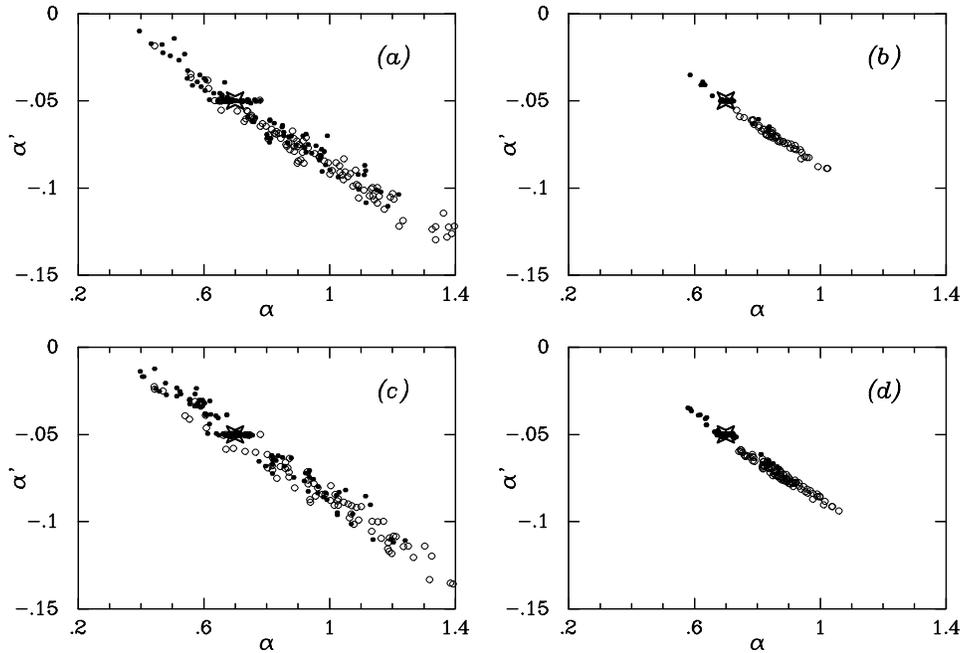}
\caption{Parameter estimates from analyses of simulated data from
a rolling power law model, using the correct (solid dots) and
incorrect (open dots) TNO survey likelihood.  Open ``$\times$'' indicates
the true parameter values.  ({\em a,b}) Estimates
using a likelihood that ignores parameter uncertainties, for
TNO samples of size 100 ({\em a}) and 1000 ({\em b}).  
({\em c,d}) Estimates
using a likelihood with incorrect source uncertainty integrals, for
TNO samples of size 100 ({\em c}) and 1000 ({\em d}).}
\end{figure*}

These simulations do not necessarily call into question the final
scientific findings reported by B04.  The simulations used a simplified
survey protocol, with somewhat larger magnitude uncertainties than B04
claim for their data, and in any case indicate that for samples with
similar size to that studied by B04, correct results are sometimes
found by chance.  What the simulations {\em do} indicate is that use of
the incorrect likelihood will eventually lead to trouble as sample
sizes get larger.

The principle lesson of this work is that {\em source uncertainties must
be carefully accounted for in analyses of survey data}.  In particular,
the effects of source uncertainties do not ``average out'' as data sets
grow in size, but in fact can grow more severe.  Bayesian inference
proves to be an ideal tool for handling this problem.  By accounting
for volumes in parameter space---especially volumes associated
with incidental parameters that
arise due to source uncertainties---a Bayesian analysis can accurately
account for the distortions introduced by source uncertainties. 
Further work on this issue, including a more thorough examination of
the B04 results, will be reported elsewhere.

%%%%%%%%%%%%%%%%%%%%%%%%%%%%%%%%%%%%%%%%%%%%%%%%
%% BACKMATTER
%%%%%%%%%%%%%%%%%%%%%%%%%%%%%%%%%%%%%%%%%%%%%%%%

\begin{theacknowledgments}
This work was supported in part by NASA AISRP grant
NAG5-12082.  My thanks to Brett Gladman and Phil Nicholson
for introducing me to the fascinating astrophysics of the
outer solar system.
\end{theacknowledgments}

%% If the bibliography is produced without BibTeX comment out the

\bibliographystyle{aipproc}   % if natbib is available

\bibliography{biblio}

\begin{thebibliography}{12}
\expandafter\ifx\csname natexlab\endcsname\relax\def\natexlab#1{#1}\fi
\providecommand{\enquote}[1]{``#1''}
\expandafter\ifx\csname url\endcsname\relax
  \def\url#1{\texttt{#1}}\fi
\expandafter\ifx\csname urlprefix\endcsname\relax\def\urlprefix{URL }\fi

\bibitem[{Jeffreys}(1938)]{Jeffreys38}
{Jeffreys}, H., \emph{Mon. Not. Roy. Ast. Soc.}, \textbf{98}, 190 (1938).

\bibitem[{Neyman} and {Scott}(1948)]{NS48}
{Neyman}, J., and {Scott}, E.~L., \emph{Econometrica}, \textbf{16}, 132 (1948).

\bibitem[{Gull}(1989)]{Gull89}
{Gull}, S., \enquote{{Bayesian Data Analysis---Straight Line Fitting},} in
  \emph{Maximum Entropy and Bayesian Methods}, edited by J.~Skilling, Kluwer,
  Dordrecht, 1989, pp. 511--518.

\bibitem[{Gladman} et~al.(2001)]{Getal01}
{Gladman}, B., {Kavelaars}, J.~J., {Petit}, J., {Morbidelli}, A., {Holman},
  M.~J., and {Loredo}, T., \emph{Astron. J.}, \textbf{122}, 1051--1066 (2001).

\bibitem[{Luu} and {Jewitt}(2002)]{LJ02}
{Luu}, J.~X., and {Jewitt}, D.~C., \emph{An. Rev. Astron. Astrophys.},
  \textbf{40}, 63--101 (2002).

\bibitem[{Gladman} et~al.(1998)]{Getal98}
{Gladman}, B., {Kavelaars}, J.~J., {Nicholson}, P.~D., {Loredo}, T.~J., and
  {Burns}, J.~A., \emph{Astron. J.}, \textbf{116}, 2042--2054 (1998).

\bibitem[{Loredo} and {Wasserman}(1995)]{LW95}
{Loredo}, T.~J., and {Wasserman}, I.~M., \emph{Astrophys. J. Supp.},
  \textbf{96}, 261--301 (1995).

\bibitem[{Loredo} and {Wasserman}(1998{\natexlab{a}})]{LW98a}
{Loredo}, T.~J., and {Wasserman}, I.~M., \emph{Astrophys. J.}, \textbf{502}, 75
  (1998{\natexlab{a}}).

\bibitem[{Loredo} and {Wasserman}(1998{\natexlab{b}})]{LW98b}
{Loredo}, T.~J., and {Wasserman}, I.~M., \emph{Astrophys. J.}, \textbf{502},
  108 (1998{\natexlab{b}}).

\bibitem[{Bernstein}(2004)]{Betal04}
{Bernstein}, G., et al., \emph{Astron. J.}, \textbf{\rm in press}, astro--ph/0308467
  (2004).

\bibitem[{Loredo} and {Lamb}(1989)]{LL89}
{Loredo}, T.~J., and {Lamb}, D.~Q., \emph{Ann. N. Y. Acad. Sci.}, \textbf{571},
  601--630 (1989).

\bibitem[{Loredo} and {Lamb}(2002)]{LL02}
{Loredo}, T.~J., and {Lamb}, D.~Q., \emph{Phys. Rev. D}, \textbf{65}, 063002
  (2002).

\end{thebibliography}

\IfFileExists{\jobname.bbl}{}
 {\typeout{}
  \typeout{******************************************}
  \typeout{** Please run "bibtex \jobname" to obtain}
  \typeout{** the bibliography and then re-run LaTeX}
  \typeout{** twice to fix the references!}
  \typeout{******************************************}
  \typeout{}
 }

\end{document}